\documentclass[%
twocolumn,
 superscriptaddress,
preprintnumbers,
nofootinbib,
 amsmath,amssymb,
 aps,
 pra,
longbibliography,
notitlepage
]{revtex4-2}

\usepackage{amsfonts}
\usepackage{graphicx}
\usepackage{color}
\usepackage{xcolor}
\usepackage[colorlinks=true, pdfstartview=FitV, linkcolor=blue, citecolor=blue, urlcolor=blue]{hyperref}

\newtheorem{theorem}{Theorem}

\newtheorem{definition}[theorem]{Definition}
\newtheorem{proposition}{Proposition}
\newtheorem{summary}{Summary}

\begin{document}
\title{Quantum gate broadcasting on graphs}
\author{Hiroki Sukeno}
\affiliation{C. N. Yang Institute for Theoretical Physics \& Department of Physics and Astronomy, State University of New York at Stony Brook, Stony Brook, NY 11794-3840, USA}
\author{Tzu-Chieh Wei}
\affiliation{C. N. Yang Institute for Theoretical Physics \& Department of Physics and Astronomy, State University of New York at Stony Brook, Stony Brook, NY 11794-3840, USA}
\begin{abstract}
Given a known or unknown phase encoded in a higher-dimensional qudit gate, it is possible to send copies of a gate that encodes the phase to multiple receivers based on a generalized quantum teleportation. 
We extend this quantum gate broadcast protocol to a quantum network on directed acyclic graphs in which agents can add phase gates to be distributed and pass unknown phase gates to subsequent receivers.
Similarly to the Greenberger-Horne-Zeilinger state, we show that the resource state can be efficiently prepared in finite time.
\end{abstract}
\date{\today}
\maketitle

\section{Introduction}

In standard quantum teleportation~\cite{bennet1993teleporting}, one can send an unknown quantum state to another party using shared entanglement and classical communication.
In quantum networks~\cite{kimble2008quantuminternet,Wehner:2018enf,cacciapuoti2018quantum,azuma2023quantumrepeaters}, on the other hand, multiple users participate in exchanging information directly or indirectly.
It is interesting to ask whether information can be sent to multiple users simultaneously using shared quantum entangled states.

The No-Cloning Theorem or No-Broadcasting Theorems~\cite{wootters1982single,dieks1982communications,barnum1996nobroadcast} impose constraints on the ability to achieve quantum broadcasting tasks.
In general, it is impossible to clone an unknown quantum state unless some compromise is made, such as allowing approximate copies~\cite{buzek1996quantum} or realizing the process as a virtual operation~\cite{parzygnat2024virtual,zheng2025experimental}.
If the state to be shared is known, one can use remote state preparation~\cite{lo2000classical,bennett2001remote,leung2003oblivious,bennett2003remote,radmark2013experimental}, which can be further simplified when the known state is restricted~\cite{agrawal2003exact,graham2015superdense,hillery2023broadcast,sukeno2023broadcast}.

In Ref.~\cite{hillery2023broadcast}, the authors considered broadcasting of restricted states with lower costs in terms of entanglement and classical communication compared to previously existing protocols.
The protocol employs one qubit per receiver and a qudit for the sender.
It was extended in Ref.~\cite{sukeno2023broadcast} to a scenario where multiple senders can collaborate in distributing copies of a phase gate to multiple receivers.
Here, we refer to the protocols in Refs.~\cite{hillery2023broadcast,sukeno2023broadcast} as \textit{quantum gate broadcasting}, as they essentially achieve the splitting of a phase gate acting on a sender's qudit into copies of a phase gate acting on the receivers' qubits.
As we will describe in the main text, in principle, the distributed phase gate does not need to be known to the sender.
This motivates our study of a quantum network in which an unknown phase gate can be routed and distributed by multiple users.

In this work, we present a quantum network in which users can add phase gates to be distributed and pass unknown phase gates to subsequent receivers.
This network is described using \textit{directed acyclic graphs}, a mathematical tool commonly used in computer science~\cite{BangJensenGutin2009}.
We define an entangled state on graphs using sequential entanglers and show that such a many-body quantum state can be efficiently prepared in finite time.
This phenomenon is analogous to that of the Greenberger-Horne-Zeilinger (GHZ) state, but our resource state exhibits several distinct features.

This paper is organized as follows.
In Section~\ref{sec:elementary}, we review our previous quantum gate broadcasting protocol.
In Section~\ref{sec:broadcasting_dag}, we present a generalization of quantum gate broadcasting to directed acyclic graphs.
In Section~\ref{sec:preparation}, we introduce an efficient preparation method for the resource state used in quantum gate broadcasting.
Finally, Section~\ref{sec:conclusion} is devoted to conclusions and discussions.

\section{Elementary quantum gate broadcasting}\label{sec:elementary}

Throughout this paper, we will consider tensor products of qudit Hilbert spaces with different dimensions. 
For a given qudit dimension $d$, We write the basis of the local Hilbert space as $\{|k \rangle\}^{d-1}_{k=0}$.
We define the shift operator as $X |k\rangle = | k+1 \,{\rm mod}\, d\rangle $, the phase operator as $Z|k\rangle = \omega^k |k\rangle$ with $\omega = e^{2 \pi i/d}$, and a parametrized phase operator as $U(\theta)|k\rangle = e^{ik\theta } |k\rangle$.

We start with reviewing the previous broadcasting protocols by us in Refs.~\cite{hillery2023broadcast,sukeno2023broadcast}.
We wish to send copies of a phase gate, i.e., $e^{i \theta Z} \otimes e^{i \theta Z}$ acting on two qubits.
If the information of the phase $\theta$ is given, one trivial way to achieve this is to prepare two pairs of Bell states and to teleport $e^{i\theta Z}$ separately through each Bell pair.
If the phase is unknown to the sender, however, it is not obvious how to achieve teleportation of such copies; this can be a situation when a phase gate is passed from someone else.
When a phase, known or unknown, is encoded in a qutrit basis, there is a way to achieve such a ``copying,'' as we review now. 

Let us consider a qutrit state $|0\rangle_A$ and a two-qubit state $|\psi\rangle_{BC}$.
We entangle the three quantum states by a generalized CX gate given by 
\begin{align}
{\rm CX}_{A \Leftarrow j} := \sum_{k=0,1}(X_A)^k   \otimes |k\rangle \langle k|_j   \quad (j=B,C) \, ,
\end{align}
where $X_A$ here is the qutrit shift gate. 
Note that the action of the qutrit shift is controlled by a qubit.
We have
\begin{align}
|\Psi\rangle = {\rm CX}_{A \Leftarrow B} {\rm CX}_{A \Leftarrow C}
\big( |0\rangle_{A} \otimes |\psi \rangle_{BC} \big) \, .
\end{align}
Then we proceed to distribute the state to Bob and Charlie via quantum channels.

After distribution of the broadcast resource state, Alice, the sender, applies a {\it qutrit} phase gate
\begin{align}
U_A(\theta) = \sum_{k=0,1,2}  e^{i\theta k}|k\rangle \langle k| \, .
\end{align}
Next, Alice measures her qutrit in the orthogonal basis 
\begin{align}
\big\{ 
Z_A^s |+\rangle_A \, \big| \, s=0,1,2 
\big\} \, ,
\end{align}
with $|+\rangle_A = \frac{1}{\sqrt{3}}( |0\rangle + |1 \rangle + |2\rangle)$ and $Z_A$ the qutrit phase operator with $d=3$.
Finally, Alice communicates her measurement result $s$ with Bob and Charlie, and they apply \begin{align}
U_B\left(\frac{2\pi s}{3}\right) \otimes U_C\left(\frac{2\pi s}{3}\right)
\end{align}
to their respective qubits, where $U_{B,C}(\theta) = \sum_{k=0,1} e^{i \theta k} |k\rangle \langle k|_{B,C}$ is the qubit phase gate.
These phase gates cancel the outcome-dependent by-product phase gates propagated from Alice's measurement.
Bob and Charlie are left with a state disentangled from Alice:
\begin{align}
U_B(\theta) \otimes U_C(\theta) |\psi\rangle_{BC}\, ,
\end{align}
where $U_{B,C}(\theta)$ is the {\it qubit} phase gate we defined above.

\begin{figure*}
\includegraphics[width=0.7\linewidth]{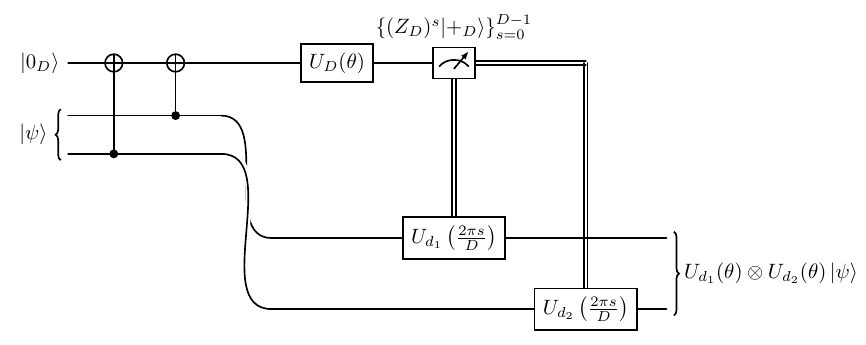}
\caption{Splitting of a phase gate into multiple receivers. 
The initial state is the tensor product of the $D$-level qudit state $|0_D\rangle$ and a general state $|\psi\rangle$ in the tensor product of $d_1$- and $d_2$-level Hilbert spaces.
A resource state is prepared with the generalized CX gates.
It is distributed over a network, followed by a $D$-level phase unitary gate encoding a parameter $\theta$. 
The sender measures their local qudit in the orthogonal basis and communicate their measurement outcome $s$ so that receivers can apply correction gates on their local qudits. 
The original phase gate $U_D(\theta)$ is then split into the tensor product of phase gates: $U_{d_1}(\theta) \otimes U_{d_2}(\theta)$.  
}
\label{fig:cascade}
\end{figure*}

This protocol is illustrated in Fig.~\ref{fig:cascade} and we refer to it as splitting of phase gates. 
It is straightforward to generalize this to a situation where Bob and Charlie have a $d_B$-level qudit and a $d_C$-level qudit, respectively, and Alice has a $d_A$-level qudit with $d_A = d_B +d_C -1$.
We refer to Proposition~\ref{prop:cascade} in Appendix~\ref{sec:formulas} for generalizations including this.

In the above, we have described a protocol in which Alice applies the gate with a phase $\theta$ known to her.
It would be more interesting to make this protocol work for an unknown phase passed from another agent through gate teleportation; this will be presented in the next section with a cascade protocol.

\section{Quantum gate broadcasting on directed acyclic graphs}\label{sec:broadcasting_dag}

It is natural to ask whether the previous gate broadcast protocol can be generalized to a larger class of networks.  
We show that the gate broadcast scheme can be generalized to {\it directed acyclic graphs} (DAG).
In particular, we devise a protocol in which one can distribute unknown quantum gates that are passed from other agents.
We call such a primitive of quantum gate broadcasting a {\it cascade}.

Let us consider a directed graph, i.e., a graph $G=(V,E)$ with vertices $V$ and arrows $E$. 
When an arrow goes from $x$ to $y$ with $x,y \in V$, we write $|x,y\rangle  \in E$. 
We call $x$ the tail of the arrow and $y$ the head of the arrow. 
DAGs are directed graphs such that, starting from a vertex, there is no path of arrows that goes back to the original vertex. 

When one can find a path that goes from $x$ to $y$, we say that $y$ is a successor of $x$, and we also say that $x$ is a predecessor of $x$; we write $y \in {\rm Succ}(x)$ and $x \in {\rm Pred}(y)$. 
A vertex without outgoing arrows is called a sink.
On the other hand, a vertex without ingoing arrows is called a source.

\subsection{Qudits on directed acyclic graphs}\label{sec:CBP_sequential_prep}

When a directed graph is a DAG, then there is a topological sorting of vertices.
That is, one can find an order of vertices $v_1v_2...v_n$ such that all the arrows are from $v_i$ to $v_j$ with $i<j$. 
Given this sorting, one can fix the ordering of the broadcast protocol.
To sinks, we assign qubits:
\begin{align}
d(v)=2 \quad {\rm for} \quad v \in {\rm sink} \, .
\end{align}
Note that it is not necessary that all sinks have $d=2$; we set this condition just for simplicity.
We give the rest of the vertices qudits with dimensions which follow a relation:
\begin{align}\label{eq:recursion}
d(v) -1 = \sum_{ |v,w\rangle \in E } \big(d(w) -1 \big) \, .
\end{align}
Namely, the local dimension of a qudit at $v$ is determined by those at vertices $w \in V$ which are direct successors of $v$.

\begin{figure}
    \includegraphics[width=0.7\linewidth]{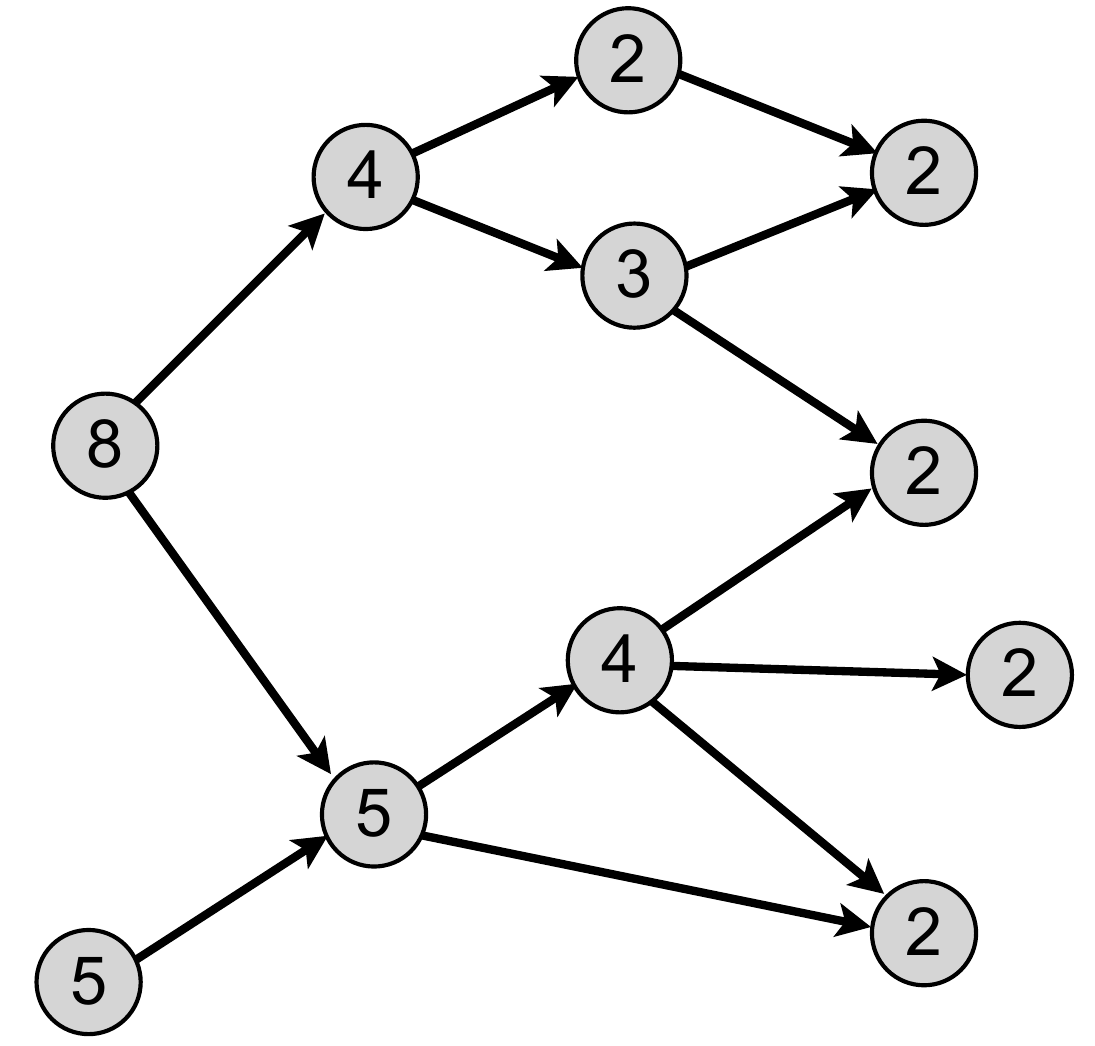}
    \caption{An example of broadcast networks on directed acyclic graphs. Numbers represent the dimension of each local qudit. Sinks are all set to be qubits. Dimensions of the rest of qudits are determined by Eq.~\eqref{eq:recursion}.}
\end{figure}

\subsection{Primitives in quantum gate broadcast}

We have given a definition of DAGs and how we assign a qudit to each vertex. 
Let us define the generalized controlled-X operator as follows:
\begin{align}
{\rm CX}_{a \Leftarrow b} := \sum_{k=0}^{d(b)-1}(X_a)^k   \otimes |k\rangle \langle k|_b  \, .
\end{align}
It applies the $d(a)$-level shift operator whose power is controlled by the $d(b)$-level state.
We also write 
\begin{align}
{\rm CX}_{b \Rightarrow a} := \sum_{k=0}^{d(b)-1}   |k\rangle \langle k|_b  \otimes ( X_a )^k \,.
\end{align}
We define the resource state for a DAG as
\begin{align} \label{eq:resource_cascade}
|\Psi(\psi)\rangle = \Big( \prod_{ |v,w\rangle \in E } {\rm CX}_{v \Leftarrow w} \Big) \cdot \Big( \bigotimes_{v \in V }| 0\rangle \bigotimes |\psi \rangle_{\rm sink} \Big)\, ,
\end{align}
where the ordering of the product follows the topological sorting of vertices: the CXs that act on sinks are placed to the right of the product.
Note that CX applies the $X$ operator on the tail of the arrow in the DAG, controlled by the head.
Hence, the directions of the arrows in the CX gates ($v \Leftarrow w$) are the opposite of those in the DAG ($ v \rightarrow w $).

\begin{figure}
    \includegraphics[width=0.7\linewidth]{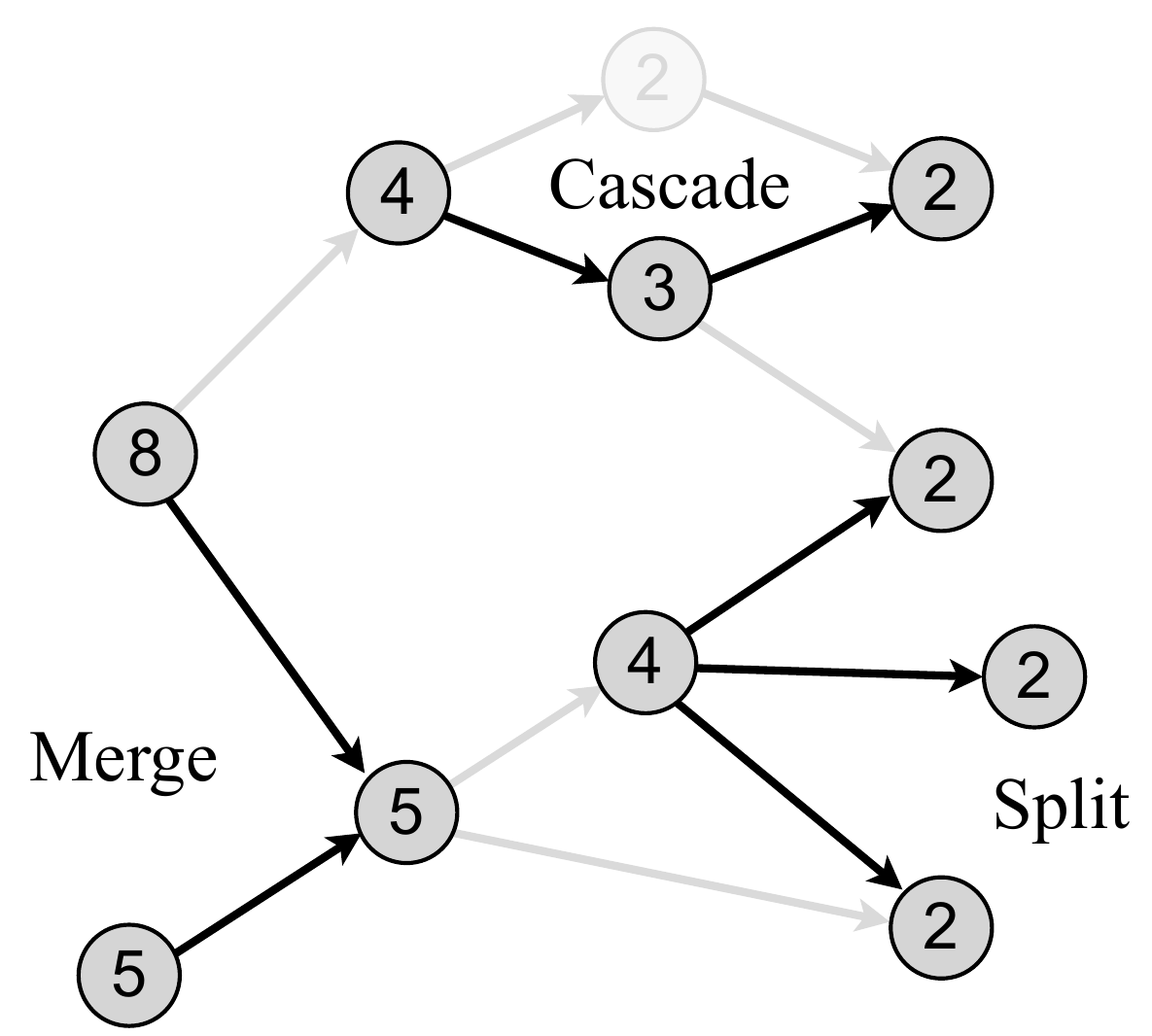}
    \caption{Examples of cascading, merging, and splitting. }
    \label{fig:primitive}
\end{figure}

In this section, we will describe primitives of cascading and merging quantum gates with small subgraphs.
The primitives are splitting, cascading, merging; see Figure~\ref{fig:primitive}.
Note that we have already explained the splitting in Section~\ref{sec:elementary} and we will not repeat it here.
Concatenating these primitives gives us a quantum gate broadcast over DAGs.

\subsubsection{Example: Cascading phase gates}

Let us consider a part of a DAG with $u,v,w \in V$ such that $|u,v\rangle , | v,w\rangle \in E$; see Figure~\ref{fig:primitive}. 
The vertex $v$ can have predecessors other than $u$ as well as successors other than $w$. 
Hence, in general, $d(u)$, $d(v)$, and $d(w)$ can be different and as $u$, $v$, and $w$ are part of a DAG,  $d(u)\ge d(v)\ge d(w)$.
Here, we omit their other predecessors and successors to focus on the relevant mechanism.

Let us consider the following state, which can be regarded as the resource state for the subgraph involving only $u$, $v$, and $w$ vertices,
\begin{align}
|\Psi_{\rm cas}\rangle 
=
\big( {\rm CX}_{u \Leftarrow v} {\rm CX}_{v \Leftarrow w}  \big)
|0\rangle_{u} \otimes |0\rangle_{v} \otimes |\psi\rangle_{w} \, .
\end{align}
Then we implement the following protocol:
\begin{enumerate}
\item One applies the $d(u)$-level phase gate $U_u(\theta_u)$ to the qudit at $u$. (One could alternatively imagine that such a phase gate is teleported from a predecessor of $u$.)
\item The agent at $u$ measures the qudit at $u$ in the qudit basis $\{ Z^{s(u)} |+\rangle \,|\, s(u)=0,...,d(u)-1 \}$.
\item The agent at $u$ communicates their measurement outcome $s(u)$ to the agent at the vertex $v$.
\item The agent at $v$ applies a correction gate $U_v\left(\frac{2 \pi s(u)}{d(u)} \right)$, where $U_v(\theta)$ is the $d(v)$-level qudit phase operator. 
\end{enumerate}
By invoking Proposition~\ref{prop:cascade} in Appendix~\ref{sec:formulas}, we see that the agents at $v$ and $w$ are left with a state disentangled from the qudit at $u$:
\begin{align}
\big( U_v(\theta_u) \otimes I \big) \cdot  
{\rm CX}_{v \Leftarrow w} \cdot \big( |0\rangle_{v} \otimes |\psi\rangle_{w} \big) \, ,
\end{align}
where $ U_v(\theta_u)$ is the $d(v)$-level phase operator acting on the qudit at $v$.
From the perspective of the agent at $v$, the gate $U_v(\theta_u)$ is unknown.

Then we continue with the following protocol:
\begin{enumerate}
\item[5.] The agent at $v$ applies the $d(v)$-level phase gate $U_v(\theta_v)$ to the qudit at $v$. 
\item[6.] The agent at $v$ measures the qudit at $v$ in the qudit basis $\{ Z^{s(v)} |+\rangle \,|\, s(v)=0,...,d(v)-1 \}$.
\item[7.] The agent at $v$ communicates their measurement outcome $s(v)$ to the agent at the vertex $w$.
\item[8.] The agent at $w$ applies a correction gate $U_w \left(\frac{2 \pi s(v)}{d(v)}\right)$, where $U_w (\theta)$ is the $d(w)$-level qudit phase operator. 
\end{enumerate}
Due to Proposition~\ref{prop:cascade} again, the agent at $w$ is left with a state disentangled from qudits at $u$ and $v$: 
\begin{align}
U_w(\theta_u + \theta_v)|\psi\rangle_{w} \, ,
\end{align}
where $ U_w(\theta_u+ \theta_v)$ is the $d(w)$-level phase operator acting on the qudit at $w$.
We also notice that the same phase gate $U_w(\theta_u+ \theta_v)$ is cascaded to successors of $v$ other than $w$. 
Hence, {\it from the perspective of the agent at $v$, they succeeded in broadcasting a phase gate unknown to themselves}. 

If the state $|\psi\rangle_w$ was actually entangled with $x$ in the form $ {\rm CX}_{w\Leftarrow x}|0\rangle_w \otimes |\psi'\rangle_{x}$ --- i.e., entangled with a further successor $x$ ---, one can imagine that this protocol can be continued where the successor $x$ receives the gate $ U_x(\theta_u+ \theta_v+\theta_w)$.
The recursive structure of the protocol implies that, for general DAGs, phase gates are accumulated through paths connected by directed arrows.

\subsubsection{Example: Merging phase gates}

Since we have demonstrated that phase gates can be split into multiple successors of a vertex, it is natural to ask whether two phase gates from two different predecessors can be merged at a common successor.
We show that this is indeed the case. 

Let us consider $v\in V$ and its predecessors $t, u \in V$ with $|u,v\rangle , | t,v\rangle \in E$.
We refer to Figure~\ref{fig:primitive} for an illustration. 

The composition of the initial state $|0\rangle_u$, the entangler ${\rm CX}_{u \Leftarrow v}$, the local gate $U_u (\theta_u )$, and the projective measurement of the qudit at $u$ forms a map acting on the qubit $v$. 
The total operation $T_{(u,v)}$ is equivalent to applying $U_v(\theta_u)$ up to a by-product phase gate and a normalization factor; see Proposition~\ref{prop:cascade}.
Similarly, the total operation $T_{(t,v)}$ between vertices $v$ and $t$ becomes a phase gate $U_v(\theta_t)$ with a by-product phase gate.
Since $T_{(t,v)}$ and $T_{(u,v)}$ commute with each other, the operation $T_{(t,v)} \circ T_{(u,v)}$ is the same as applying $U_v(\theta_u + \theta_t)$ up to by-product phase gates.

Hence, phase gates from two vertices can be merged at a common successor. 
Clearly, one can have as many predecessors as they wish.

\subsection{Quantum gate broadcasting on directed acyclic graphs}

\begin{figure}
    \includegraphics[width=0.9\linewidth]{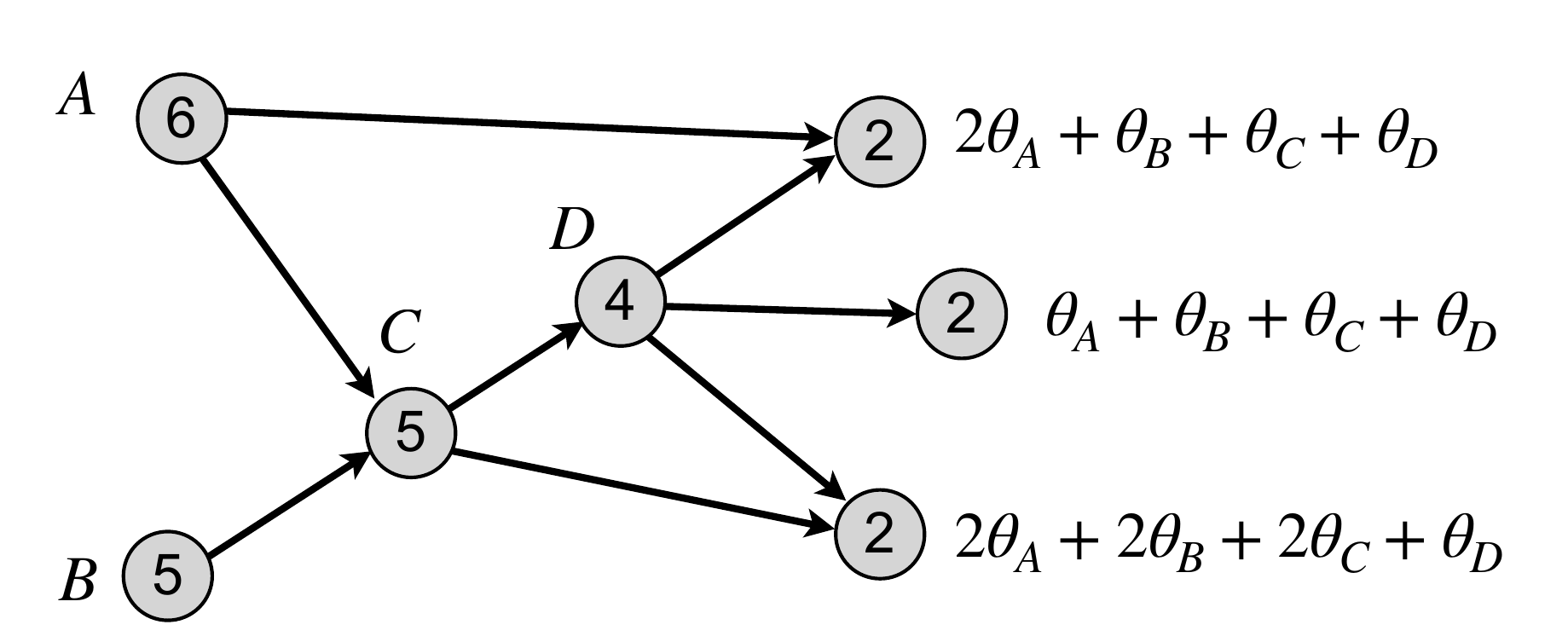}
    \caption{An example of quantum gate broadcast networks. The three sinks receive phases from other vertices after the protocol.}
    \label{fig:phase_sum}
\end{figure}

Combining the above primitives --- cascading, splitting, and merging of phase gates --- gives us a quantum gate broadcasting scheme on DAGs. 
We have the following result.
\begin{summary}[Quantum gate broadcasting]
With the distributed resource state given in Eq.~\eqref{eq:resource_cascade}, each agent assigned to the vertex $v$ performs: (i) applying $d(v)$-level phase gate $U_v(\theta_v)$; (ii) applying the $d(v)$-level phase gate $\prod_{|u,v\rangle \in E} U_v \left(\frac{2 \pi s(u)}{d(u)} \right)$ based on the measurement results of direct predecessors $\left\{u \,\, {\rm s.t.}\,\, |u,v\rangle \in E \right\}$; (iii) measuring the qudit in the orthogonal basis; and (iv) communicating their measurement outcome to successors via classical channels.  

Then, the qubit state associated with sinks of the DAG becomes 
\begin{align}
\Big(\bigotimes_{v \in {\rm sink}} U_v(\Theta_v)\Big) |\psi \rangle \, , 
\end{align}
with $\Theta_v = \sum_{w \in {\rm Pred} (v)} n_{(v,w)} \theta_w$. The integer $n_{(v,w)}$ is the number of independent paths that connect from $w$ to $v$ in the DAG.
\end{summary}
We give an illustration of an example in Figure~\ref{fig:phase_sum}.

\section{Efficient preparation schemes}\label{sec:preparation}

We analyze preparation schemes for the quantum gate broadcast resource $|\Psi(\psi)\rangle$ with the wave function $|\psi\rangle$ at sink qubits; see Eq.~\eqref{eq:resource_cascade}.
In the previous section, we gave a definition of the state based on initial disentangled states and sequential (generalized) CX gates. 
We first present an efficient scheme for preparing the Greenberger-Horne-Zeilinger state on a one-dimensional open chain. 
The method here is an adaptation of techniques developed in recent studies~\cite{piroli2021quantum,tantivasadakarn2024long-range, lu2022measurement,bravyi2022adaptive,tantivasadakarn2023hierarchy,li2023symmetry, okuda2024anomalyCSS,ren2024efficient}, where efficient preparation schemes have been constructed for long-range entangled states such as Calderbank-Shor-Steane codes and solvable anyon models.
We generalize this to the preparation of the quantum gate broadcasting resource state.

\subsection{A motivating example}

Let us consider a one-dimensional open chain.
We aim to produce the following state, which is a special case of Eq.~\eqref{eq:resource_cascade} with no branching:
\begin{align}
|\Psi (\psi) \rangle 
&= \bigg( \prod^{N-1}_{j=1} {\rm CX}_{j \Leftarrow j+1} \bigg)
\otimes^{N-1}_{k=1}|0 \rangle_k \otimes |\psi\rangle_{N} \nonumber \\
&\propto
\bigg( \prod^{N-1}_{j=1} \frac{1+Z_{j}Z_{j+1}}{2} \bigg)
\otimes^{N-1}_{k=1}|+ \rangle_k \otimes |\psi\rangle_{N}
\, .
\end{align}
The proportionality is understood by first seeing that both expressions satisfy $Z_j Z_{j+1} |\Psi (\psi) \rangle = |\Psi (\psi) \rangle$ for $j=1,..., N-1$, and this set of conditions leaves only the one qubit state $|\psi \rangle$ unfixed.
Namely, the wave function follows 
\begin{align}
k_j - k_{j+1} = 0 \, \qquad (1 \leq j\leq N-1) \, ,
\end{align}
with $k_j$ the bit $\{0,1\}$ of the $j$-th qubit in the computational-basis expansion of the wave function.  
When the boundary state is chosen as $|\psi\rangle = |+\rangle$, the state is the Greenberger-Horne-Zeilinger state.
The bulk product state $\otimes^{N-1}_{k=1}|+ \rangle_k$ is chosen so that, when we choose $|\psi\rangle = |+\rangle$, the two computational bases $\{|0...0\rangle , |1...1\rangle\}$ appear in the unbiased manner with plus signs.
(In the language of the stabilizer circuit~\cite{gottesman1998heisenberg}, we set the initial stabilizers $\{+X_j\}^{N-1}_{j=1}$ so that (i) they all anti-commute with the $Z_jZ_{j+1}$ projections; and (ii) for the Greenberger-Horne-Zeilinger state we have $+\prod_j X_j$ as a correct stabilizer after the projections.)

To generate the state above, one can construct a product state on integer sites $\otimes^{N-1}_{k=1}|+ \rangle_k$ and $|\psi\rangle_N$ to begin with. 
Then we introduce a copy of the $|+\rangle$ product state at half integers $h$ such that $1 < h < N$. 
Next, we apply a product of controlled-Z gates, so we have
\begin{align}
&\bigg( \prod_{j \in \{1,...,N\}} {\rm CZ}_{j, j+\frac{1}{2}} {\rm CZ}_{j, j-\frac{1}{2}}  \bigg)  \nonumber \\
& \qquad \cdot \bigotimes_{k} |+\rangle_{k+\frac{1}{2}} 
\bigotimes_{k} |+\rangle_{k}
\bigotimes |\psi\rangle_N
\, .
\end{align}
We measure the half-integer sites in the $X$ basis. 
The sequence of the $|+\rangle_{x+\frac{1}{2}}$ state, adjacent CZ gates, and the $X$-basis measurement at $x+\frac{1}{2}$ gives us a projector enforcing $Z_x Z_{x+1} = \pm 1$.

Let us assume that we obtained $|-\rangle_{x+\frac{1}{2}}$ as the outcome at $x+\frac{1}{2}$. The output state on the integer sites will have a sign-flipped stabilizer:
\begin{align}
Z_x Z_{x+1} |\Psi_{\rm output}\rangle = -  |\Psi_{\rm output}\rangle \, , 
\end{align}
which means
\begin{align}
|\Psi_{\rm output}\rangle &\propto 
\bigg( \frac{1- Z_{x}Z_{x+1}}{2} \prod_{j\neq x} \frac{1+Z_{j}Z_{j+1}}{2} \bigg) \nonumber \\
& \cdot \Big( \otimes^{N-1}_{k=1}|+ \rangle_k \otimes |\psi\rangle_{N} \Big) \, ,
\end{align}
assuming that all the other outcomes are $|+\rangle$. 
One can remedy this flipped outcome by applying a string of $X$ operators starting at $j=1$ and ending at $j=x$, i.e., $S_x= \prod^x_{j=1}X_j$.
We get 
\begin{align}
S_x |\Psi_{\rm output}\rangle = |\Psi (\psi)\rangle \, .
\end{align}
Clearly, this correction procedure can be generalized to arbitrary patterns of measurement outcomes by taking a product of string operators for each flipped outcome.

\subsection{Efficient preparation of quantum gate broadcast resource states}

We generalize the measurement-based efficient preparation scheme for the Greenberger-Horne-Zeilinger state to the resource state for the quantum gate broadcasting on DAGs.
Let $\sigma$ be the set of sinks in a DAG. 
We write the set of vertices without sinks as $\widetilde{V} = V \backslash \sigma$. 
In the resource state, the following equation is satisfied:
\begin{align} \label{eq:k_relation}
{\sf K}_v := k_v
- \sum_{|v,w\rangle \in E} k_{w} = 0 \qquad v \in \widetilde{V} \, ,
\end{align}
where $k_v$ is the integer $0 \leq k_v \leq d(v)-1$ of the qudit at $v$ in the qudit computational-basis expansion of the wave function. 
To see this, one can use Proposition~\ref{prop:CNOT_comm} to confirm that 
\begin{align}
&W_v := \left[ U_v\left( \frac{2 \pi}{d(v)} \right) \prod_{|v,w\rangle \in E} U_w \left(\frac{-2 \pi}{d(v)} \right) \right]  \, , \\
& W_v |\Psi (\psi)\rangle = |\Psi (\psi)  \rangle   \label{eq:W_psi_psi}
\end{align}
for $ v \in \widetilde{V}$. 
The phase operator $W_v$ on the left hand side reads as follows on the qudit computational basis:
\begin{align}
\exp\left[ 
\frac{2\pi i}{d(v)} {\sf K}_v 
\right] \, .
\end{align}
Since $0 \leq k_v \leq d(v)-1$ for $v \in \widetilde{V}$ and $d(v)-1 = \sum_{|v,w\rangle \in E} (d(w)-1)$ by definition, we have that $ -(d(v)-1)\leq {\sf K}_v  \leq d(v)-1$.
Therefore, the condition~\eqref{eq:W_psi_psi} is equivalent to the condition~\eqref{eq:k_relation}.

The resource state $|\Psi(\psi)\rangle$ is the state in which $W_v =+1$ is obeyed for $v \in \widetilde{V}$ and qudit computational bases appear without bias up to that from $\psi$. 
Let us write the projector that enforces $W_v =+1$ as $\mathbb{P}_v$. 
We have that
\begin{align}
|\Psi (\psi) \rangle \propto \bigg( \prod_{v \in \widetilde{V}} \mathbb{P}_v \bigg)\cdot \bigg(\bigotimes_{v \in \widetilde{V}} |+\rangle_v \otimes |\psi\rangle_{\sigma}\bigg) \, .
\end{align}
The projectors $\{ \mathbb{P}_v\}_{v \in \widetilde{V}}$ commute with one another as they all act diagonally on the computational basis: They can be applied simultaneously.
Hence, our final goal is to realize each projector as an efficient physical operation.

We assign an additional vertex $v'$ to $v\in \widetilde{V}$ and introduce an ancillary $d(v)$-level qudit $|+\rangle_{v'}$ to it.
Let ${\rm CW}_{v',v}$ be the following controlled-phase gate:
\begin{align} \label{eq:cw}
{\rm CW}_{v',v} = \sum_{ \ell = 0}^{d(v)-1} |\ell\rangle \langle \ell |_{v'} \otimes (W_v)^\ell \, . 
\end{align}
We consider measurement of the qudit $v'$ in the $X$ basis. 
When we obtain the $|+\rangle_{v'}$ as the measurement output, the sequence of operations acts as the projector:
\begin{align}
\langle +|_{v'} {\rm CW}_{v',v} |+\rangle_{v'}
= \frac{1}{d(v)} \sum_{\ell = 0}^{d(v)-1}  (W_v)^\ell   = \mathbb{P}_v.
\end{align}
To see this more concretely, one can apply the operator on the left hand side of the equations on the computational basis $| \{k_v\}_{v\in V}\rangle$ and find that it acts as  
\begin{align}
\frac{1}{d(v)}\sum^{d(v)-1}_{\ell=0} \exp\left[ \frac{2 \pi i}{d(v)} \ell \cdot {\sf K}_v\right] \, ,
\end{align}
which is zero unless ${\sf K}_v=0$.
This is the ideal projector that yields the resource state. 
In general, if we obtain $Z^{s_v} |+\rangle$ ($s_v = 0, ..., d(v)-1$) as the measurement output, the sequence of operations will enforce ${\sf K}_v = s_v$.
This can be remedied --- without ruining ${\sf K}_v=0$ conditions for other vertices --- by applying the product of shift gates over all vertices that contain $v$ as a successor:
\begin{align}
S[s_v] = \prod_{ u \in {\rm Pred}(v) } (X_u)^{n_{(u,v)}s_v} \, . 
\end{align}
Here, $n_{(u,v)}$ is the number of independent paths that connect from $u$ to $v$ in the DAG.
Then we get
\begin{align}
S[s_v] | \Psi_{\rm output} \rangle = | \Psi (\psi) \rangle \, . 
\end{align}
This correction procedure can be generalized to arbitrary patterns of measurement outcomes by taking a product of $\{ S[s_v] \}_{v \in \widetilde{V}}$:
\begin{align}\label{eq:total_correction}
S[\{s_v\}] = \prod_{v \in \widetilde{V}} S[s_v] \, . 
\end{align}

\begin{figure*}
    \centering
    \includegraphics[width=0.8\linewidth]{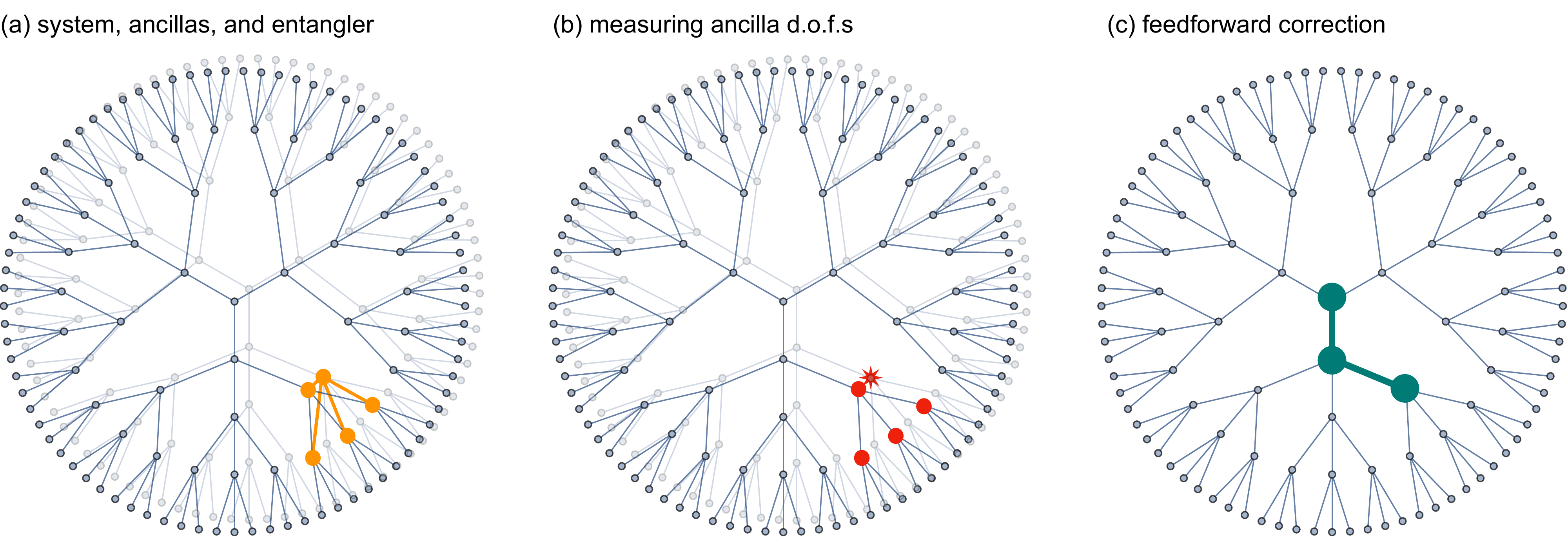}
    \caption{Finite-time preparation protocol. 
    Our protocol works for general directed acyclic graphs. Here we illustrate it for a tree graph for simplicity.
    (a) Given a graph, we introduce a copy to each vertex, where we place an ancilla qudit. We then entangle the ancilla qudit with the adjacent qudits on the main graph. (b) We measure the ancilla qudits. The red star represents the place where non-ideal measurement outcome occurred. (c) We erase the effect of non-ideal measurement outcome at $v$ by applying a string operator. It acts on all the vertices which are predecessors of $v$.}  
    \label{fig:prep_const_time}
\end{figure*}

We illustrate the preparation protocol for the tree graph in Figure~\ref{fig:prep_const_time}.
We summarize our scheme as follows.
\begin{summary}[Efficient preparation]
The resource state~\eqref{eq:resource_cascade} for the phase-gate broadcasting protocol on DAGs can be prepared as follows: (i) prepare the product of $|\psi\rangle_\sigma$ ($\sigma$: sinks), $\bigotimes_{v \in \widetilde{V}}|+\rangle_{v}$ ($\widetilde{V}$: the set of vertices with sinks removed), and $\bigotimes_{v \in \widetilde{V}}|+\rangle_{v'}$ ($v'$: a copy of $v$); (ii) apply the entangler~\eqref{eq:cw}; (iii) measure each qudit at $v'$ in the qudit $X$ basis; and (iv) apply the feedforward operation~\eqref{eq:total_correction}.
\end{summary}

\subsection{Removing edges from graphs}\label{sec:detaching_subgraph}

The above preparation protocol utilizes a state
\begin{align}\label{eq:state_prep}
    &|\Psi_{\rm prep}(\psi)\rangle \nonumber \\
    &\quad = \bigg( \prod_{v \in \widetilde{V}} {\rm CW}_{v',v} \bigg) \cdot \bigg( \bigotimes_{v \in \widetilde{V}}  
    |+\rangle_{v'} \otimes  
    |+\rangle_{v}   \bigg)
    \otimes |\psi\rangle_\sigma \, 
\end{align}
before measuring $v'$ qudits in the $X$ basis.
Here, we notice analogy to the graph state~\cite{hein2004multiparty,hein2006entanglement}.
In particular, if we instead measure a vertex $v'$ in the $Z$ basis, the entangling gate ${\rm CW}_{v',v}$ acts as the $W_v$ gate to some power that depends on the measurement outcome, and this does not induce entanglement to the post-measurement state. 
Therefore, we see that {\it if one measures a vertex $v'$ of $|\Psi_{\rm prep}(\psi)\rangle$ in the $Z$ basis, edges $\{ |v,w\rangle  \, | \, w \in V   \}$ will be removed from the quantum broadcasting network.}
In other words, there will be no entanglement across such edges, and phase gates will not cascade through them.
A particularly interesting case would be the tree graph, where a $Z$ measurement of a vertex $v'$ will detach branches of $v$ from the network completely.

We propose to use this as a feature for agents to remove some of the successors from the network. 
To utilize this, instead of the state~\eqref{eq:resource_cascade}, the state~\eqref{eq:state_prep} is distributed to all agents, each of whom will possess both $v$ and $v'$.
Our quantum network is hence dynamical.

\section{Conclusions and Discussion}\label{sec:conclusion}

We have presented a scheme in which multiple agents can participate in adding phase gates to be distributed and passing unknown phase gates to subsequent receivers.
Phase gates can be split, merged, and cascaded over the quantum network.
The resource state used for this quantum network can be prepared in finite time.
Agents can remove edges from the network graph.  

In our previous work~\cite{hillery2023broadcast,sukeno2023broadcast}, we discussed applications of broadcast protocols, such as quantum key distribution~\cite{phoenix2000three} and authentication~\cite{dutta2022review}.
It would be interesting to explore a multi-party version of quantum key distribution as a potential application of the protocol introduced in this work.
We also note that transmission of phase gates finds applications in private quantum computation~\cite{broadbent2009universal,fitzsimons2017private}.
Another potential application lies in interferometry, similar to Refs.~\cite{gottesman2012longerbaseline, khabiboulline2019optical}.
It would be worth investigating whether cascading in our quantum gate broadcast can be used to amplify signals in such setups.

There is room for further theoretical exploration of our protocol.
First, can general directed graphs with cycles be used?
It would be interesting to determine whether gate broadcasting functions with such a resource state and whether it offers any advantages.
On another front, we also examined whether the cascade resource could be used for the backward propagation of information.
So far, we have not found capability of backward broadcasting in the cascade broadcast resource state.
It would be intriguing to explore whether any quantum entangled state allows for two-way quantum gate broadcasting.
We leave these as open questions.

The distributed quantum phase gates are implemented on the edge state as the result of measurement of `bulk' vertices in a graph.
From a broader perspective, our setup is reminiscent of the measurement-based quantum computation~\cite{raussendorf2001one,raussendorf2003measurement,briegel2009measurement,wei2018quantum}, where resource states are typically (symmetry-protected) topologically ordered, and logical gates act on the edge mode~\cite{pollmann2012symmetry}; see Refs.~\cite{gross2007novel,gross2007measurement,miyake2010quantum,sukeno2023measurement,wong2024gauge}.
Our construction may offer an interesting arena for both quantum network problems and condensed matter physics.

\section*{Acknowledgement}
The authors acknowledge useful discussions with Shuyu Zhang, Himanshu Gupta and CR Ramakrishnan.
This work was supported by the National Science Foundation through Grant No. FET-2106447.
The authors also acknowledge the support from the Center for Distributed Quantum Processing at Stony Brook University.

\bibliography{ref}

\appendix
\section{Formulas}\label{sec:formulas}

\begin{definition}
Let $\{ |k \rangle \}^{d-1}_{k=0}$ be the basis of a $d$-dimensional Hilbert space. 
We define the qudit Pauli operators as follows:
\begin{align}
X &= \sum^{d-1}_{k=0}|k +1 \,\,{\rm mod }\,\, d \rangle \langle k| \, , \\
Z &= \sum^{d-1}_{k=0} \omega^k |k \rangle \langle k| \, ,
\end{align}
where $\omega = e^{2 \pi i /d} $.
We also define a phase gate:
\begin{align}
U(\theta) = \sum^{D-1}_{k=0} e^{i \theta k} | k \rangle \langle k | \, .
\end{align}
Namely, $Z \equiv U\left(\frac{2 \pi}{d}\right)$.
\end{definition}
We give the following set of definitions.
\begin{definition}
Let $d_j \geq 2$ be a positive integer and $\mathcal{H}_{j}$ be a $d_j$-level qudit Hilbert space with  $j \in \{1,...,N\}$. 
Let $\mathcal{H}_A$ be a $D$-level qudit system with $D= 1+ \sum^{N}_{j=1} (d_j-1)$. 
We define a CX gate $\mathcal{H}_j \times \mathcal{H}_A \rightarrow \mathcal{H}_j \times \mathcal{H}_A$ as follows:
\begin{align}
\label{eq:appCNOT}
{\rm CX}_{j\Rightarrow A} = \sum^{d_j -1}_{k=0} | k \rangle \langle k |_j \otimes (X_A)^k \, .
\end{align}
When acting on $\mathcal{H}_1 \times \cdots \times \mathcal{H}_N \times \mathcal{H}_A$, it acts as the identity operator except for $\mathcal{H}_j$ and $\mathcal{H}_A$.
\end{definition}

\begin{proposition}
$X$, $Z$, $U(\theta)$, and CX are unitary operators.
\end{proposition}

\begin{definition}
We define the qudit plus state as $|+\rangle := \frac{1}{\sqrt{d}} \sum^{d-1}_{k=0} |k \rangle $. 
We also define $|\widetilde{s}\rangle = Z^s |+\rangle$. 
\end{definition}
\begin{proposition}
Both $\{ |s \rangle \langle s|  \}^{d-1}_{k=0}$ and $\{ |\widetilde{s} \rangle \langle \widetilde{s}|  \}^{d-1}_{k=0}$ form an orthogonal projection, i.e., a measurement basis.
\end{proposition}

\begin{proposition}\label{prop:CNOT_comm}
For the $D$-level Hilbert space $\mathcal{H}_A$, the $d_j$-level Hilbert space $\mathcal{H}_j$, and the unitary operators we defined above, we have the following identity for operators on $\mathcal{H}_j \times \mathcal{H}_A$:
\begin{align}
&\bigg[ I_j \otimes U_A(\theta) \bigg]\cdot {\rm CX}_{j\Rightarrow A} \nonumber \\
&\qquad = {\rm CX}_{j \Rightarrow A} \cdot\bigg[ U_j(\theta) \otimes U_A(\theta) \bigg] \, , 
\end{align}
where subscripts $j$ and $A$ denote that the operators act on $\mathcal{H}_j$ and $\mathcal{H}_A$, respectively.
In particular, for $U_A\left(\frac{2 \pi}{D} \right) \equiv Z_A$, we have
\begin{align} 
& \bigg[ I_j \otimes (Z_A)^\ell \bigg] \cdot {\rm CX}_{j \Rightarrow A} \nonumber \\
&\qquad = {\rm CX}_{j\Rightarrow A} \cdot \bigg[ U_j\left(\frac{2 \pi \ell}{D}\right) \otimes (Z_A)^\ell \bigg] \, .
\end{align}
\end{proposition}

\begin{proposition}[Splitting phase gate]\label{prop:cascade}
The following identity for operators $\mathcal{H}_1 \times \cdots \times \mathcal{H}_N \rightarrow \mathcal{H}_1 \times \cdots \times \mathcal{H}_N$ holds:
\begin{align}
&\langle \widetilde{s}|_A  
U_A(\theta)  \bigg(\prod^{N}_{j=1} {\rm CX}_{j \Rightarrow A} \bigg)  | 0\rangle_A  \nonumber \\  
&\qquad = 
\frac{1}{\sqrt{D}}\bigotimes^{N}_{j=1} \left[ U_j \left(\frac{-2 \pi s}{D}\right) \cdot  U_j(\theta) \right]  
\, . 
\end{align}
\end{proposition}

\end{document}